\documentclass[twocolumn,prb,showpacs]{revtex4}%
\usepackage{graphicx}%
\usepackage{amsmath}%
\usepackage{amsfonts}%
\usepackage{amssymb}
\usepackage{bm}

\def\s{{\sigma}}
\def\e{{\epsilon}}
\def\k{{ {\bm k} }}
\def\p{{ {\bm p} }}

\def\w{{\omega}}
\def\a{{\alpha}}
\def\b{{\beta}}

\begin{document}
\title{
Impurity-Induced Electronic Nematic State and \\
$C_2$-Symmetric Nanostructures in Iron-Pnictide Superconductors
}
\author{Yoshio \textsc{Inoue}$^{1}$,
Youichi \textsc{Yamakawa}$^{1}$, and
Hiroshi \textsc{Kontani}$^{1}$
}
\date{\today }

\begin{abstract}
We propose that impurity-induced electronic nematic state
is realized above the orthorhombic structure transition temperature $T_S$
in iron-pnictide superconductors.
In the presence of strong orbital fluctuations near $T_S$,
it is theoretically revealed that a single impurity induces local
orbital order with $C_{2}$-symmetry,
consistently with recent STM/STS measurements.
Each impurity-induced $C_2$-symmetric nanostructure
aligns along $a$-axis
by applying tiny uniaxial pressure along $b$-axis.
In this impurity-induced nematic phase, 
the resistivity shows sizable in-plane anisotropy
($\rho_b/\rho_a\sim2$) even above $T_S$,
actually observed in various ``detwinned'' samples.
The present study indicates the existence of strong
orbital fluctuations in iron-pnictide superconductors.
\end{abstract}

\address{
$^1$ Department of Physics, Nagoya University and JST, TRIP, 
Furo-cho, Nagoya 464-8602, Japan. 
}
 
\pacs{74.70.Xa, 74.20.-z, 74.20.Rp}

\sloppy

\maketitle


\section{Introduction}

Since the discovery of iron-pnictide superconductors,
\cite{Hosono},
a lot of effort has been devoted to understand the 
overall phase diagram,
including the superconducting (SC) state in the tetragonal (T) phase
and non-SC orthorhombic (O) phase.
In Ba(Fe,Co)$_2$As$_2$, the O structure transition at $T_S$ is second-order 
\cite{second-order}, 
and very large softening of shear modulus $C_S$ suggests 
the existence of strong ferro-quadrupole $\phi_S\propto {\hat x}^2-{\hat y}^2$ 
fluctuations above $T_S$
\cite{Fer,Yoshizawa-old,Kontani-soft,Goto}.
In many compounds, the superconducting (SC) transition temperature
$T_{\rm c}$ takes the highest value
near the endpoint of the O phase,
suggesting a close relation between the superconductivity 
and the orbital instability.
The weak ferro-quadrupole order in the O phase induces
the spin-density-wave with ${\bm Q}=(\pi,0)$
\cite{Kontani-soft}.

As for the SC mechanism,
spin fluctuation mediated sign-reversing $s$-wave state 
($s_\pm$-wave state) had been proposed from the early stage,
noticing on the Coulomb interaction and intra-orbital nesting
between hole- and electron-pockets
\cite{Kuroki,Mazin,Hirschfeld}.
In iron-pnictides, each Fermi pockets are mainly composed of 
$t_{2g}$ orbitals of Fe atoms.
On the other hand, orbital fluctuation mediated
$s$-wave state without sign reversal ($s_{++}$-wave state) 
had been investigated in Refs. \cite{Kontani,Saito,Onari-FLEX}:
Strong orbital fluctuations originate from
the inter-orbital nesting between Fermi pockets
in the presence of Coulomb and weak electron-phonon ($e$-ph) interactions.
The latter scenario is supported by the robustness of the 
SC state against impurities in many iron-pnictides
\cite{Sato-imp,Nakajima,Li,Onari-impurity},
and by the orbital independent SC gap observed
in Ba122 systems by laser ARPES measurement
\cite{Shimo-Science,Saito}
Also, experimental ``resonance-like'' hump structure 
in the neutron inelastic scattering is well reproduced in terms 
of the $s_{++}$-wave SC state, rather than the $s_\pm$-wave SC state,
by taking the suppression in the inelastic scattering
in the SC state (dissipationless mechanism)
\cite{Onari-resonance}.

According to Ref. \cite{Kontani-soft},
the structure transition originates from the
ferro-charge quadrupole $\phi_S=O_{x^2-y^2}\propto n_{xz}-n_{yz}$ instability,
realized by the bound state formation of two orbitons with opposite momenta.
By this two-orbiton theory, we can fit the temperature dependence of $C_S$ 
in Ba(Fe$_{1-x}$Co$_x$)$_2$As$_2$ for $x=0\sim0.16$ almost perfectly 
\cite{Kontani-comment}.
The spin nematic theory (or two-magnon process)
 \cite{Fer,Kievelson} is another candidate.
However, incommensurate spin order is realized 
in Ba(Fe$_{1-x}$Co$_x$)$_2$As$_2$ for $x\ge0.056$ \cite{incomme},
although the latter theory requires commensurate fluctuations.


Furthermore, recent discovery of ``electronic nematic transition'' 
in the T phase, free from any lattice deformation, has been 
attracting great attention.
For example, in ``detwinned'' Ba(Fe$_{1-x}$Co$_x$)$_2$As$_2$ 
\cite{detwin,detwin-rev}
under very small uniaxial pressure ($\sim5$MPa), 
sizable in-plane anisotropy of resistivity 
emerges at $T^*$, which is about 10K$\sim$100K higher than $T_S$.
The nematic order is also observed in BaFe$_2$(As,P)$_2$
by the magnetic torque measurement 
\cite{Matsuda}.
Now, it is demanded to find
which degree of freedom, spin or orbital, is more important for 
the nematicity, orthorhombicity and superconductivity.



In this paper, we discuss the impurity-induced electronic nematic phase
in iron-pnictides, using the mean-field approximation (MFA) in real space.
When orbital fluctuations develop,
we obtain various types of local orbital orders with lower symmetries
($C_4$, $C_{\rm 2v}$, $C_2$, etc.),
actually reported by STM/STS 
autocorrelation analyses \cite{Davis,Song}.
The large cross section of the local order 
gives giant residual resistivity,
far beyond the 
$s$-wave unitary scattering value; $\sim20\mu\Omega$cm/\%.
When $C_2$ nanostructures are aligned along $a$-axis,
the in-plane anisotropy of resistivity reaches $40$\%,
consistently with experiments \cite{detwin,detwin-rev}.
Such large anisotropy is not achieved when
isotropic impurity scattering is considered
\cite{Chen,Fer2}.

In annealed Ba(Fe$_{1-x}M_x$)$_2$As$_2$ ($M$=Co, Ni),
the difference $|\rho_b-\rho_a|$ 
is very small in the absence of $M$-impurities ($x=0$) \cite{Uchida},
while it increases in proportional to $x$ for $x\le4$\%
 \cite{detwin-rev,Uchida}.
In contrast, both the magnetic moment and lattice deformation 
monotonically decrease with $x$. 
These facts strongly support the idea of impurity-induced 
nanostructures.

In strongly correlated electron systems,
impurity potential frequently causes drastic change 
in the electronic state.
For example, in nearly antiferromagnetic metals, magnetic 
correlation is extremely enhanced near the nonmagnetic impurity site,
giving rise to the local magnetic moment ($\sim1\mu_{\rm B}$)
and large residual resistivity
 \cite{Kontani-review}
that are indeed observed in optimally- and under-doped cuprates.
In terms of weak-coupling scheme,
such phenomena originate from the Friedel oscillation
since the large local-density-of-states (LDOS) sites 
could trigger the strong fluctuations around the impurity.
As for the iron-based superconductors,
the system would be close to antiferro-orbital critical point.
Thus, it is natural to expect the occurrence 
of ``impurity-induced local orbital order'' in iron pnictides.

\section{Model Hamiltonian and Method of Calculation}

Here, we study the single-impurity problem due to 
orbital-diagonal impurity potential $I$ \cite{Kontani}
in a large cluster with $800$ Fe sites,
based on the two-dimensional ten-orbital tight-binding model for 
LaFeAsO in Refs. \cite{Kuroki,Miyake}.
We set $x$ and $y$ axes parallel to the nearest Fe-Fe bonds.
Then, the Fermi surfaces are mainly composed of $t_{2g}$ orbitals 
($xz$, $yz$ and $xy$),
although $e_g$ orbitals also play non-negligible roles.
Here, we consider the following quadrupole-quadrupole interaction
\cite{Kontani,Saito,Onari-FLEX,Kontani-soft}: 
\begin{eqnarray}
H_{\rm quad}=-g\sum_{i}\left\{ {\hat O}_{xz}^i {\hat O}_{xz}^i
+{\hat O}_{yz}^i {\hat O}_{yz}^i+{\hat O}_{xy}^i {\hat O}_{xy}^i\right\}
\label{eqn:Hquad}
\end{eqnarray}
where ${\hat O}_{\Gamma}^i$ is the quadrupole operator 
for channel $\Gamma$ at site $i$ introduced in Ref.\cite{Kontani-soft}:
${\hat O}_{\Gamma}^i =\sum_{l,m,\s}o_{\Gamma}^{l,m}c_{i,l\s}^\dagger c_{i,m\s}$,
where $o_{\Gamma}^{l,m}$ is the matrix element of the charge quadrupole operator.
Note that ${\hat O}_{\mu\nu}\propto {\hat l}_\mu{\hat l}_\nu
+{\hat l}_\nu{\hat l}_\mu$.
The quadrupole coupling constant $g$ in eq. (\ref{eqn:Hquad})
originates from both the $e$-ph interaction as well as the 
Coulomb interaction for the charge sector,
as discussed in Refs. \cite{Kontani,Saito}.
Since we are interested in the nonmagnetic orbital order,
we neglect the Coulomb interaction to simplify the calculation.
Then, strong orbital fluctuations for $\Gamma=xz,yz$ channels are produced 
by relatively small $g$ ($\sim0.2$ eV)
\cite{Kontani,Saito,Onari-FLEX}.
Hereafter, the unit of energy is eV.

Here, we put $T=0.02$ and the electron filling $n=6.0$ per Fe,
which corresponds to undoped compounds like BaFe$_2$As$_2$.
In the absence of impurity, the bulk antiferro-orbital order 
occurs for $g>g_{\rm c}\equiv0.222$.
Below, we study the following mean-field equation for $g<g_c$:
\begin{eqnarray}
M_{l,m}^i= \langle c_{i,l\s}^\dagger c_{i,m\s}\rangle_{I,g}
-\langle c_{i,l\s}^\dagger c_{i,m\s}\rangle_{I,0}
\label{eqn:M}
\end{eqnarray}
where $i$ is the Fe site, 
and $l,m$ represent the $d$-orbital.
$M_{l,m}^i$ is impurity-induced mean-field; 
${\hat M}^i=0$ for $I=0$.
Then, the mean-field potential due to Hartree term is 
\begin{eqnarray}
S_{l,m}^{i}= \sum_{l',m'}\Gamma^c_{lm,l'm'}M_{l',m'}^i
\label{eqn:Sigma}
\end{eqnarray}
where $\Gamma_{lm,l'm'}^c=-2g \sum_\Gamma^{xz,yz,xy}o_\Gamma^{lm}o_\Gamma^{l'm'}$
is the bare interaction for charge sector \cite{Kontani-soft}, 
and the mean-field Hamiltonian is
${\hat H}_{\rm MF}={\hat H}_0+\sum_i {\hat S}^i+ {\rm const}$.
In the MFA, we solve eqs. (\ref{eqn:M})-(\ref{eqn:Sigma}) self-consistently.

\section{Numerical Results and Discussions}

\begin{figure}[!htb]
\includegraphics[width=.95\linewidth]{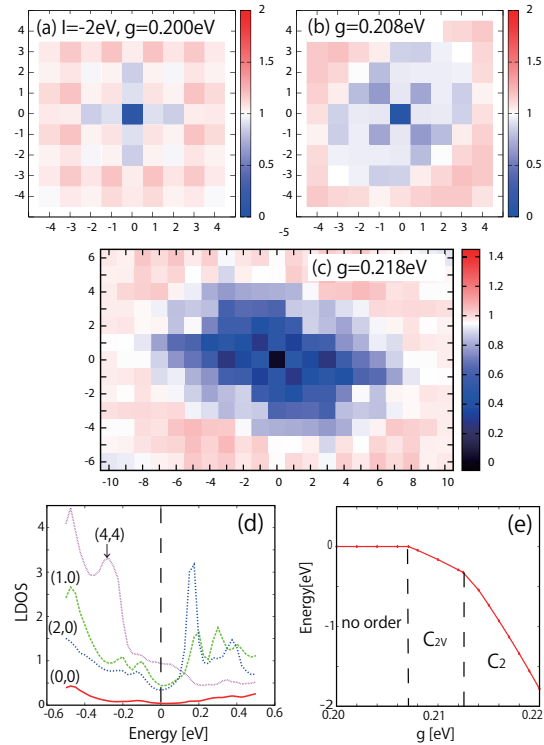}
\caption{(color online)
Obtained LDOS at $E_{\rm F}$ for $I=-2$ and 
(a) $g=0.200$: without orbital order,
(b) $g=0.208$: orbital order with diagonal $C_{\rm 2v}$-symmetry, and
(c) $g=0.218$: orbital with $C_{2}$-symmetry.
(d) Energy-dependence of the LDOS for $g=0.218$.
(e) $g$-dependence of the free-energy.
}
\label{fig:LDOS}
\end{figure}

In Figs. \ref{fig:LDOS} (a)-(c), we show the obtained DOS 
at Fermi level ($E_{\rm F}$) in real space,
in which the center is the impurity site with $I=-2$.
For $g=0.200$ (a), the impurity-induced mean-field is absent.
The small modulation of the LDOS around the impurity
is caused by the Friedel oscillation.
For $g>0.207$, impurity-induced local orbital order with diagonal 
$C_{\rm 2v}$ symmetry appears, as shown in (b).
The suppression of the DOS is caused by the orbital order,
consistently with a recent optical conductivity measurement \cite{optical}.
With increasing $g$, the orbital order
changes to $C_2$ symmetry for $g>0.212$, shown in (c).
The size of the nanostructure is $\sim15 a_{\rm Fe-Fe}$
($\sim7 a_{\rm Fe-Fe}$) along $x$ ($y$) axis.
Such a large impurity-induced object is actually observed
in Ba(Fe,Co)$_2$As$_2$ by STM/STS \cite{Davis,Song}.
When the impurity concentration $n_{\rm imp}$ is $\sim1$\%,
the obtained $C_2$-order would be stabilized by the weak
overlap between neighbors 
against thermal fluctuations omitted in the MFA.
(Similar $C_2$-order is also realized for $I=\infty$.)
Figure \ref{fig:LDOS} (d) shows the energy-dependence of LDOS 
for $g=0.218$ at ${\bm r}=(0,0)$ (impurity site), $(1,0)$, $(2,0)$, 
and $(4,4)$.
Near $(0,0)$, the LDOS is modified for a wide energy range.
Figure \ref{fig:LDOS} (e) presents the free-energy as function of $g$.
In the MFA, each transition at $g\approx 0.207$ and $0.212$
is first-order.

\begin{figure}[!htb]
\includegraphics[width=.75\linewidth]{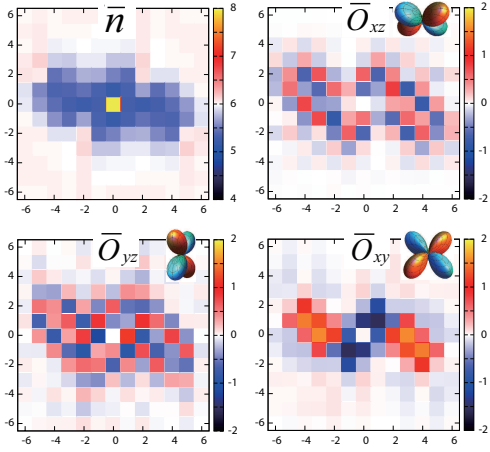}
\caption{(color online)
Obtained electron density ${\bar n}^i$ and
quadrupole order ${\bar O}_{\Gamma}^i$ at Fe sites
for $I=-2$ and $g=0.218$:
}
\label{fig:MF}
\end{figure}

Since $M_{l,m}^i=M_{m,l}^i$,
the present mean-field has 15 components at each site.
They are represented as charge density or monopole ($l=0$),
quadrupole ($l=2$), and hexadecapole ($l=4$) orders.
The first two orders are give as
${\bar n}^i= 2\sum_{l,l} M_{l,l}^i$ and
${\bar O}_{\Gamma}^i= 2\sum_{l,m} o_{\Gamma}^{l,m}M_{l,m}^i$,
%
%
where $\Gamma=xz$, $yz$, $xy$, $z^2$, and $x^2-y^2$.
The hexadecapole order is negligibly small in the present study.
Figure \ref{fig:MF} shows the dominant four mean-fields,
${\bar n}^i$, ${\bar O}_{xz}^i$, ${\bar O}_{yz}^i$ and ${\bar O}_{xy}^i$,
for $g=0.218$.
We verified that the quadrupole interactions for $\Gamma=xz/yz$ channels
in eq. (\ref{eqn:Hquad}) are indispensable for the $C_2$-order.
The obtained quadrupole order is very difference from the 
uniform quadrupole ordered state (${\bar O}_{x^2-y^2}\propto n_{xz}-n_{yz}=$const.)
in the orthorhombic phase \cite{Kontani-soft},
and therefore the impurity-induced nematic order will exist even below $T_S$.

The $C_2$-order in Fig. \ref{fig:LDOS} (c) can be aligned by 
the strain-induced quadrupole potential;
$H'= \Delta E \sum_i {\hat O}_{x^2-y^2}^i$ and 
$\Delta E =\eta_S\e_S\cdot \chi_{x^2-y^2}^Q({\bm 0})/\chi_{x^2-y^2}^{(0)}({\bm 0})$,
where $\e_S\propto a-b$ is the strain and $\eta_S$ is the 
strain-quadrupole coupling.
$\chi_{x^2-y^2}^Q({\bm 0})$ is the ferro-quadrupole susceptibility,
which is strongly enhanced near $T_S$ due to the two-orbiton process
as discussed in Ref. \cite{Kontani-soft}.
This would be the reason why the nematic ordered state
is easily detwinned by small uniaxial pressure near $T_S$.
In fact, detwinning by uniaxial pressure is possible
only when the structure transition is the second-order \cite{Proz}.

\begin{figure}[!htb]
\includegraphics[width=.9\linewidth]{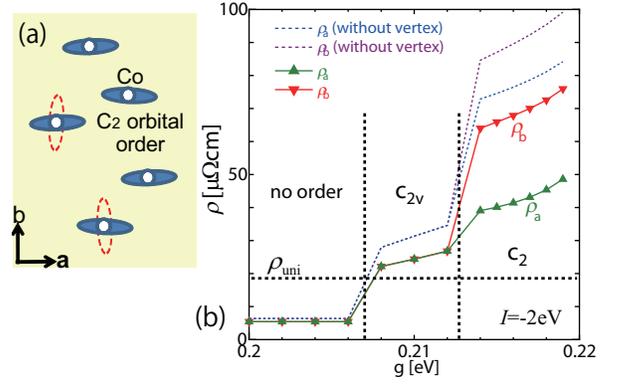}
\caption{(color online)
(a) Alignment of the impurity-induced $C_2$-orders
under uniaxial pressure ($a>b$).
(b) Obtained $\rho_{a(b)}$ for $n_{\rm imp}=1$\% and $I=-2$:
$\rho_b>\rho_a$ in the nematic phase.
}
\label{fig:rho-Im2}
\end{figure}

Here, we assume $x\parallel a$-axis and $y\parallel b$-axis.
In detwinned compounds with $a>b$,
ARPES measurements indicates $\Delta E<0$, 
{\it i.e.}, $n_{xz}>n_{yz}$ \cite{ARPES2,ARPES3}.
For a single $C_2$-order, we obtain the relation 
$F_a-F_b\approx 2.5\Delta E$, where $F_{a(b)}$ is the 
free-energy when the $C_2$-order is along $a(b)$-axis.
Therefore, the nematic order along $a$-axis is realized 
by detwinning ($a>b$), schematically shown in Fig. \ref{fig:rho-Im2} (a).
Note that two kinds of $C_2$-orders, 
the $C_2$-order in Fig \ref{fig:LDOS} (c)
and its inversion with respect to $x$ axis,
still degenerate and coexist with equal probability.

Now, we calculate the in-plane resistivity 
in the nematic state shown in \ref{fig:rho-Im2} (a).
We use the $T$-matrix approximation, which gives the exact result
when $n_{\rm imp}\ll1$ and localization is negligible.
The $T$-matrix is given by solving the following equation
in the orbital-diagonal basis:
\begin{eqnarray}
{\hat T}_{{\bm r},{\bm r}'}(\w)&=&({\hat I}+{\hat S})_{\bm r}
\delta_{{\bm r},{\bm r}'}
 \nonumber \\
& &+ \sum_{{\bm r}''}({\hat I}+{\hat S})_{\bm r}
{\hat G}^{(0)}_{{\bm r}-{\bm r}''}(\w){\hat T}_{{\bm r}'',{\bm r}'}(\w)
 \label{eqn:T}
\end{eqnarray}
where ${\hat S}_{\bm r}$ is the impurity-induced mean-field potential,
and ${\hat G}^{(0)}_{\bm r}(\w)$ is the Green function without impurities.
${\hat I}_{\bm r}=I{\hat 1}\delta_{{\bm r},{\bm 0}}$ is the impurity potential
\cite{Kontani}.
The $T$-matrix is non-local when ${\hat S}_{\bm r}\ne0$.
After the Fourier transformation, the self-energy 
in the $T$-matrix approximation is
${\hat \Sigma}(\k,\w)= n_{\rm imp}{\hat T}_{\k,\k}(\w)$,
and the full Green function is
${\hat G}(\k,\w)=(\w+\mu-{\hat H}_\k^0-{\hat \Sigma}(\k,\w))^{-1}$.
Then, the in-plane conductivity is given as 
\begin{eqnarray}
\sigma_\nu=\frac{e^2}{\pi}\frac1N\sum_{\k,\a}
{v}_{\k,\nu}^\a{J}_{\k,\nu}^\a|G_\a(\k,i\delta)|^2
\end{eqnarray}
where $\nu=x$ or $y$, and $\a$ represents the $\a$th band.
${v}_{\k,\nu}^\a$ is the group velocity 
and $G_\a(\k,\w)$ is the full Green function in the band-diagonal basis.
${J}_{\k,\nu}^\a$ is the total current including the vertex correction,
which is given by solving the following Bethe-Salpeter equation:
$\displaystyle J_{\k,\nu}^\a = v_{\k,\nu}^\a
 +\frac1N\sum_{\p,\b} I_{\k,\p}^{\a,\b}|G_\b(\p,i\delta)|^2 J_{\p,\nu}^\b$
%
where $I_{\k,\k'}^{\a,\b}=n_{\rm imp}|T^{\a,\b}_{\k,\k'}(i\delta)|^2$ 
is the irreducible vertex.

The obtained results for $I=-2$ and $n_{\rm imp}=1$\%
are shown in Fig. \ref{fig:rho-Im2} (b).
Here, we assume the inter-layer distance is 0.6nm.
Without orbital order, the resistivity is $5.5\mu\Omega$cm,
which is about one-fourth of the maximum value without orbital order:
$\rho_{\rm uni}\sim 20\mu\Omega$cm for $I\approx +1$.
When diagonal $C_{2v}$-order appears, the resistivity exceeds $\rho_{\rm uni}$,
due to large cross section of the ``effective impurity radius''
as recognized in Fig. \ref{fig:LDOS} (b).
In the nematic phase with horizontal $C_{2}$-order,
we obtain large anisotropy $\rho_b/\rho_a\sim 2$:
By including the vertex correction,
both $\rho_a$ and $\rho_b$ are suppressed 
and the anisotropy $\rho_b/\rho_a$ is enlarged,
since the contribution of the forward scattering 
is correctly subtracted.
The averaged resistivity $(\rho_a+\rho_b)/2$
per 1\% impurity reaches $\sim50\mu\Omega$cm,
which is comparable to the residual resistivity by 1\% Co
impurities observed in La1111 \cite{Sato-imp} and
Ba122 \cite{Uchida}.

Now, we discuss the nematic transition at $T^*$ in real compounds.
Beyond the MFA,
the effective interaction ${\tilde g}\ (<g)$ decreases with $T$
due to the thermal fluctuation  \cite{Onari-FLEX}.
Then, one possibility is that 
the phase transition from the diagonal $C_{\rm 2v}$ to vertical $C_2$ 
occurs at $T^*$.
(Then, ${\tilde g}\approx 0.212$ at $T^*$.)
Other possibility is that $C_2$ order is realized even above $T^*$,
while the necessary condition for detwinning, $\chi_{x^2-y^2}^Q({\bm 0})\gg1$,
is satisfied only below $T^*$.
In both cases, experimental results can be explained.


\begin{figure}[!htb]
\includegraphics[width=.8\linewidth]{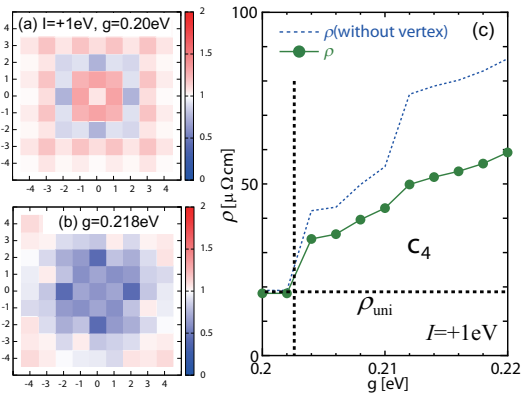}
\caption{(color online)
Obtained LDOS at $E_{\rm F}$ in the case of $I=+1$ and
(a) $g=0.200$: without orbital order, and
(b) $g=0.218$: orbital order with $C_4$ symmetry.
(c) Obtained resistivity for 1\% impurity with $I=+1$.
}
\label{fig:I1}
\end{figure}

We also study the impurity-induced local orbital order for $I=+1$.
Figure \ref{fig:I1} (a) shows the LDOS without orbital order:
The realized Friedel oscillation pattern different from Fig. \ref{fig:LDOS} (a)
would induce a new type of orbital order.
In fact, we obtain the orbital order with $C_4$ symmetry for $g>0.203$:
Figure \ref{fig:I1} (b) shows the LDOS for $g=0.218$.
We also obtain a meta-stable solution with $C_2$ symmetry
similar to Fig. \ref{fig:LDOS} (c),
whose free-energy is about $0.1$ eV higher than that
for the $C_4$ symmetry solution.
(When $I=-2$, the 
$C_4$ symmetry solution is ``unstable'' with positive free-energy.)
Figure \ref{fig:I1} (c) shows the resistivity $\rho=\rho_a=\rho_b$
for $I=+1$ and $n_{\rm imp}=1$\%:
It exceeds the unitary value as soon as $C_4$-order appears,
and it reaches $\sim50\mu\Omega$cm for $g\sim g_{\rm c}$.

It is noteworthy that the obtained $C_4$ order
looks similar to Sn-impurity-induced ``ring-shape object'' in LiFeAs
observed by Hanaguri \cite{Hana} very recently.
The realized large reduction in the DOS would result in the 
suppression of the $s_{++}$-wave state.
In fact, in BaFe$_{1.89-2x}$Zn$_{2x}$Co$_{0.11}$As$_2$, the suppression in 
$T_{\rm c}$ per 1\% Zn-impurity is $-\Delta T_{\rm c}/\% \sim3$ K/\%
 \cite{Li}:
Such small suppression of $T_{\rm c}$ 
is consistent with the $s_{++}$-wave state,
since $-\Delta T_{\rm c}/\% \sim 20$ K/\% is expected in the $s_\pm$-wave state
when the mass-enhancement is $m^*/m_b \sim 3$
 \cite{Onari-impurity}.

Finally, we consider the impurities on other than Fe sites.
We expect that impurity-induced nematic phase is realized
in BaFe$_2$(As,P)$_2$, since $P$ sites 
give finite impurity potential on the neighboring four Fe sites.
In this case, we actually obtain impurity-induced order 
with $C_{2}$- or $C_{1h}$-symmetry, 
consistently with experiments \cite{Kontani-comment}.
In contrast, nematic state is not realized in (K,Ba)Fe$_2$As$_2$
\cite{no-nematic},
maybe because K sites are outside of FeAs planes.

\section{Summary}

In summary, we discussed
impurity-induced electronic nematic state
based on the orbital fluctuation theory.
The obtained local orbital orders with various symmetries
($C_{\rm 2v}$, $C_2$, and $C_4$) are consistent with recent STM/STS measurements.
In the case of $C_2$-order, the anisotropy 
of resistivity reaches $\rho_b/\rho_a\sim2$, which presents
a natural explanation for the nematic state 
in various ``detwinned'' iron-pnictides.
Thus, characteristic features of iron pnictides,
nematic and structure transitions as well as superconductivity, 
are well understood based on the orbital fluctuation theory.


\acknowledgements
We thank Y. Matsuda, T. Shibauchi, S. Uchida, H. Eisaki, M. Sato, 
M. Itoh, Y. Kobayashi, T. Hanaguri, D. Hirashima, S. Onari and T. Saito
for valuable discussions.
This study has been supported by Grants-in-Aid 
from MEXT of Japan, and by JST, TRIP.


\end{document}